\documentstyle[12pt,aas2pp4,flushrt]{article}
\tightenlines
\def\kms{\relax \ifmmode {\,\rm km\,s}^{-1}\else \,km\,s$^{-1}$\fi}

\def\farcs{\hbox{$.\!\!^{\prime\prime}$}}

\def\secd#1.#2{ #1\farcs#2 }               % seconds over decimal point

\def\gr{$^\circ$}

\def\nii{[N~{\sc ii}]}
\def\sii{[S~{\sc ii}]}

\def\oi{[O~{\sc i}]}

\def\oiii{[O~{\sc iii}]}

\lefthead{Low-ionization structures in PNe}
\righthead{Gon\c calves et al.}

\begin{document}

\title{Low-ionization structures in planetary nebulae: 
       confronting models with observations}

\author{Denise R. Gon\c calves}
        
\affil{Instituto de Astrof\'{\i}sica de Canarias, c. Via L\'actea S/N, \\
	E--38200 La Laguna, Tenerife, Spain \\e--mail: denise@ll.iac.es}

\author{Romano L. M. Corradi}

\affil{Isaac Newton Group of Telescopes, Apartado de Correos 321, E-38700, \\
       Sta. Cruz de la Palma, Spain \\e--mail: rcorradi@ing.iac.es}

\author{Antonio Mampaso}
        
\affil{Instituto de Astrof\'{\i}sica de Canarias, c. Via L\'actea S/N,\\
       E--38200 La Laguna, Tenerife, Spain \\e--mail: amr@ll.iac.es}

\begin{abstract}

Around 50 planetary nebulae (PNe) are presently known to possess
``small-scale" low-ionization structures 
(LISs) located inside or outside
their main nebular bodies. We consider here the different kinds of LISs
(jets, jet-like systems, symmetrical and non-symmetrical knots) and
present a detailed comparison of the existing model predictions with the 
observational morphological and kinematical properties.

We find that nebulae with LISs appear indistinctly spread among all
morphological classes of PNe, indicating that the processes leading to 
the formation of LISs are not necessarily related to those responsible 
for the asphericity of the large-scale morphological components of PNe.

We show that both the observed velocities and locations of most
non-symmetrical systems of LISs can be reasonably well reproduced
assuming either fossil condensations originated in the AGB wind or
in situ instabilities.

The jet models proposed to date (hydrodynamical and magnetohydrodynamical
interacting winds or accretion-disk collimated winds) appear unable
to account simultaneously for several key characteristics of the
observed high-velocity jets, such as their kinematical ages and the
angle between the jet and the symmetry axes of the nebulae. 
The linear increase in velocity observed in several jets  
favors  magnetohydrodynamical confinement  compared to
pure hydrodynamical interacting wind models.

On the other hand, we find that the formation of jet-like systems 
characterized by
relatively low expansion velocities (similar to those of the main
shells of  PNe) cannot be explained by any of the existing models. 

Finally, the knots which appear in symmetrical and opposite pairs of 
low velocity could be understood as the survival of fossil (symmetrical) 
condensations formed during the AGB phase or as structures that have 
experienced substantial slowing down by the ambient medium. 
\end{abstract}


\section{Introduction}

The most accepted scenario for the formation of a planetary nebula
(PN) is that originally proposed by Kwok, Purton \& Fitzgerald
(1978), the so called interacting stellar wind (ISW) model.   ISW
models are successful at predicting the formation and  properties of
the main morphological components of PNe (Kwok 1994; Mellema 1995),
such as bright rims, attached shells and the haloes that characterize
spherical and elliptical PNe. These main structures are better
identified in the light of hydrogen and helium recombination lines, as
well as in the forbidden \oiii\ lines.  But, on usually smaller
scales, there are also structures which are instead more prominent in
low-ionization lines, such as [N~{\sc ii}], [O~{\sc ii}], \oi\ and \sii. 
They have
often been grouped under the general denomination of ``small-scale
low-ionization structures'' of PNe.  These low-ionization structures
(hereafter we use the acronym LISs \footnote{The acronymous LIS is used
here only for the sake of conciseness, with no intention of adding a
new term to the already very rich nomenclature used for PNe (see
Schwarz 2000).})  have received more and more attention after the
imaging catalogues of Balick (1987), Schwarz, Corradi \& Melnick
(1992), Manchado et al.  (1996), G\'orny et al.  (1999) and the
compilation by Corradi et al. (1996), the latter particularly devoted
to low-ionization structures.
 
Observationally, LISs appear with a variety of morphologies: knots,
tails, filaments, jets and jet-like structures of low ionization
attached to, or detached from, the main shells of the nebulae. They
are sometimes labeled with specific acronyms intending to describe
some of their physical characteristics -- for instance, FLIERs (fast,
low-ionization emission regions; Balick et al.  1993); or BRETs
(bipolar, rotating, episodic jets; L\'opez, V\'azquez, \& Rodr\'\i guez
1995).

On the theoretical side, different models have been proposed to
explain the origin of LISs.  The main ones are: ISW models (Frank,
Balick, \& Livio 1996; Garc\'\i a-Segura et al.  1999); jet formation
and its interaction with the circumstellar medium (Cliffe et al.
1995; Redman \& Dyson 1999), and the interaction of shells with the 
interstellar medium (Soker \& Zucker 1997). In addition,
other ingredients -- such as stellar magnetic fields, rotation,
precession, a binary system in the center, and dynamical
(Kelvin--Helmholtz and Rayleigh--Taylor) and/or radiation instabilities
-- are sometimes considered within these models in order to explain
the observations. In spite of this theoretical effort, the nature of
LISs in PNe is still poorly understood.

In the following sections, we present a comprehensive compilation of
the observations of LISs in PNe (focusing on morphological and
kinematical data) and discuss their different properties with the aim
of confronting the various theoretical models proposed to explain
their formation.  The motivation for this study is presented in
Section 2. In Section 3, we present the PNe sample. Section 4 is
devoted to jet and jet-like LISs, Section 5 to symmetrical knots, and
Section 6 to non-symmetrical LISs. A final discussion on LISs showing
multiple pairs of LIS is presented in Section 7, and the main
conclusions are summarized in Section 8.

\section{Motivation}

Low-ionization small-scale structures potentially contain important 
information about the mass loss and radiative processes that lead to
the formation and development of a planetary nebula.  Recently, we
have obtained morphological and kinematical data for nine 
PNe (Corradi et al. 1997, 1999, 2000a, 2000b) with the aim of
studying the physical properties, origin and evolution of the LISs
therein contained. Results are both promising and
puzzling since we have found LISs with notably different properties
relative to each other, in terms of expansion velocities, shapes,
sizes and locations relative to the main nebular components. It
appears that several physical processes have to be considered in order
to account for the formation and evolution of all the different LISs
that we observed.

In particular, the following basic questions are open. Are magnetic
fields -- either in single or binary stars -- 
necessary for producing jets in PNe, as generally believed for young 
stellar objects and active galactic nuclei jets?  Which processes, 
even in complex systems
like interacting binary stars, can produce multiple systems of highly
collimated outflows expanding in directions almost perpendicular to
 each other as observed in some PNe? How do low-velocity 
collimated LISs form?  Are symmetric pairs of knots recent ejecta from
the central stars, or fossil condensations tracing a peculiar 
mass-loss geometry during the AGB?  Are non-symmetrical LISs formed by in
situ instabilities?

Taking advantage of our recent work and of the information which
is spread throughout the literature, we carefully address these points 
in what follows.  

\section{The sample of PNe with LISs}

\subsection{Morphological/kinematical classification for LISs}

In this paper, we adopt a working definition of LISs as features especially
prominent in low-ionization lines (the most commonly observed is \nii), and
which have a size, at least in one direction, much smaller than the main
morphological components of the PNe, namely the main shells and haloes of
elliptical PNe, or the lobes of bipolar PNe.  In spite of this definition, it
should be noted that, in some cases, LISs can form large structures which can
extend  to few parsecs, such as the string of knots which define the
point-symmetrical collimated outflow of Fg~1 (Palmer et al. 1996).

Due to the variety of LISs found, some further definition of
the terms used in the following is in order. Hereafter we refer to as {\it
knots} all unresolved LISs, as well as all resolved small-scale low-ionization
features with an aspect ratio (maximum length to maximum width) close to
1. All features with an aspect ratio much larger than one are instead called
{\it filaments}. Knots and filaments appear with any orientation with respect
to the central star, and not necessarily in pairs.

{\it Jets} are a more restrict subclass of highly collimated filaments,
which i) are directed in the radial direction from the central star, ii)
appear in opposite symmetrical pairs and iii) move with velocities
substantially larger than those of the ambient gas which form the main bodies
of the nebulae. Finally, all features resembling jets,
but for which no evidence exists that are expanding significantly more
rapidly than
the ambient gas (in many cases, because of lack of information and/or
appropriate modeling) are called {\it jet-like} structures. It is clear
that projection effects, which are often poorly known, play a fundamental
role in distinguishing genuine jets from jet-like LISs.

In our classification, ansae (Balick 1987) and FLIERs (Balick et al. 1993)
are pairs of knots, whereas BRETs (L\'opez et al. 1995) are pairs of knots or
jets, with a point-symme\-trical distribution. 

With these definitions in mind, and
aiming at a more comprehensive analysis of the properties of LISs than 
has been 
attempted in the past, we have added to our observational sample all the data
that we were able to recover from the literature, building up a final sample
of 50~PNe containing LISs. These are listed in Table~1, together with the 
designations
for the PNe and the morphology of the main nebular shell. The fourth column 
in Table 1 contains the classification of the LISs. The references from which 
the properties of the nebulae were taken are given in the last column of the 
Table.

\subsection{LISs frequency vs. PN morphological type}

In the third column of Table~1, we give the morphological classification of
the PNe following the scheme of Schwarz, Corradi \& Stanghellini (1993),
based on optical narrow-band images. Extended PNe are classified into four main
classes: elliptical (E), bipolar (B), point-symmetric (P) and irregular (I).

In our sample, we find 29 elliptical PNe ($58\% $), seven bipolars ($14\%$), 
nine point-sym\-metrics ($18\%$) and three irregulars ($6\%$). Two PNe 
 (K~4-47 and IC~2149) were not 
classified since they do not easily fit into any of 
the above classes.  These figures can be compared with those for the 
global sample of PNe. Corradi \& Schwarz (1995) classified 359 PNe, finding 
64\% ellipticals, 14\% bipolars, 4\% point-symmetrics, and 18\%  
irregulars. Thus, LISs appear indistinctly in elliptical and bipolar 
objects compared to the general sample of PNe, but seem to be more frequent in 
point-symmetrical objects. The latter result, however, is probably 
related to the definition itself of this class of PNe, whose characteristic 
symmetry is generally defined by the presence of small-scale structures. 
Irregulars are much less frequent in our sample than in the general one, 
probably because new observations allow a better classification of these 
objects.

We therefore conclude that low-ionization structures are spread throughout
the morphological classes of PNe indistinctly. Such a result would indicate
that the formation of LISs is not necessarily connected with the processes
responsible for the asphericity in PNe (whatever they are). 

\section{Jets and jet-like LISs}

\subsection{Theories for the formation of collimated LISs}

To allow for a critical discussion of the most relevant observational
properties of jets in PNe, a brief review of  jet formation models is
presented here.

Interacting stellar wind (ISW) theories, in addition to explaining the
formation of the main morphological components of PNe, can, under certain
conditions, account for the formation of highly collimated LISs. The first ISW
model for jet formation (Icke et al.  1992) was able to produce moderately
collimated structures by inertial confinement starting from a torus-like
density contrast in the mass distribution of the AGB wind.  Subsequently, it
became clear that the transition from the slow to the fast wind (the so
called momentum-driven phase in the early evolution of a PN)  also plays an
important role in the confinement (cf.  Frank et al. 1996).
Nowadays, ISW models consider in detail the evolution of the fast wind in
velocity and mass loss (Dwarkadas \& Balick 1998), resulting in considerably
more structure on smaller scales than in the models in which the fast wind
velocity is assumed to be constant. In addition to  jet formation, the
momentum-driven phase leads to instabilities (thin-shell Vishniac
instabilities and Kelvin--Helmholtz instabilities) which may be the origin of
knots and filaments. Later in the evolution, during the 
energy-driven phase, other instabilities, such as Rayleigh--Taylor ones, can
also appear, modifying the structures formed in the previous phase
(Breitschwerdt \& Kahn 1990; Dwarkadas \& Balick 1998; Garc\'\i a-Segura et
al. 1999).  The above ISW models, based on a single-star scenario for  
PN/jet formation, consist of hydrodynamical (HD)
simulations (Mellema \& Frank 1997; Borkowski, Blondin, \& Harrington 1997;
in addition to the papers cited above) or consider the presence of a
magnetized stellar wind (MHD simulations; 
R\'o$\dot{\rm z}$yczka \& Franco 1996; Garc\'\i a-Segura
1997; Garc\'\i a-Segura et al. 1999).

As commented previously, a key parameter for  HD collimation is the
equator-to-pole density contrast of the AGB wind. Even if  HD ISW
models are able to account for the jet formation, the question of
the source of that asphericity in the AGB wind remains open; the most
popular models consider that the central star is part of a binary
system (Morris 1987; Mastrodemos \& Morris 1999; see also Soker \&
Livio 1994 and Livio \& Pringle 1997, for the formation of 
accretion-disks around the PN central star). In this way an accretion-disk and
possibly also an ``excretion'' disk can be formed and later used for
collimating the jet, as well as for shaping the PNe. On the other hand,
in the MHD ISW models presently available, the non-spherical density
distribution in the AGB wind is the direct result of stellar rotation,
which, together with the magnetic tension, produces aspherical shells
and highly collimated jets.

The ISW simulations specifically predict certain important
properties that can be compared with the observations. Firstly, the
jets formed i) are two-sided highly collimated structures, ii) possess
supersonic velocities much larger than those of the main shells, iii)
are roughly coeval with the main shell, and iv) lie along the major
symmetry axis of the PNe (unless precession of the center star is considered,
in which case point-symmetric jets are formed). Secondly, when a
magnetic field is taken into account, the expansion velocity of the
jets is expected to increase linearly outwards (Garc\'\i a-Segura et
al. 1999).  Finally, depending on the kind of collimation
mechanism assumed, the jets may appear only outside the main shell
(pure HD confinement) or also inside it (MHD collimation).

As an alternative to the standard ISW theory, some models consider the
possibility that high-velocity, highly collimated jets in PNe are
directly produced by the accretion-disks, which result from the
interaction between the central star of a PN (or its progenitor) with
a secondary.  Most studies consider that the jet forms immediately
after the common-envelope phase of a close binary system  
(Soker 1992, 1996; Soker \& Livio 1994; Reyes-Ruiz \& L\'opez
1999). However, the evolution from the common-envelope phase up to the
jet formation has not been investigated in detail. Mastrodemos \&
Morris (1998, 1999) consider the formation of accretion-disks in
long-period binaries, thereby avoiding the common-envelope phase.

In a different context, it seems clear now that the formation of
highly collimated, powerful jets, both in young stellar objects and in
active galactic nuclei, requires two main ingredients: an accretion-disk 
and a vertical magnetic field (Livio 1997, 2000). Very recently,
Blackman et al. (2000) proposed an MHD model for the interplay of
stellar and disk winds which constrains the stellar and disk rotation
and the magnetic fields in order to account for multiple bipolar
outflows, either bipolar lobes or highly collimated jets.

In general, jets emerging from accretion-disks arise before\footnote{A
few peculiar, iron-deficient post-AGB stars in close binaries show
circumstellar disks fuelled by back-accretion during the post-AGB
evolution (Van Winckel et al. 1999). In principle, such disks could
produce jets {\it after} the formation of the PN main shell.} the
formation of the PN main shell. Typical velocities are several
hundred \kms, whereas their orientation depends on the angle between
stellar and disk rotational (and magnetic) axes, i.e., they are not
necessarily aligned with the nebular symmetry axis (Blackman et al.
2000). Depending on the stage of evolution of the primary, these
accretion-disk jets are expected to be found outside or inside the
bright rims.

Finally, precession of jets within 3D simulations (Cliffe et al. 1995;
Garc\'\i a-Segura 1997) is the natural way to account for the
point-symmetry in PNe and in their LISs. Note that Cliffe et al. (1995)
describe the interaction of a precessing jet with the circumstellar
medium, regardless of the jet formation mechanism.

\subsection{Observational properties of jets}

We present in Table 2 the list of the 12 known low-ionization jets.
In this table, column 2 gives a (subjective) confidence index which
reflects the quality of the available observational data and
spatio-kinematical modeling performed. Their approximate kinematical
ages (compared to that of the main nebular shell), the angle between
the jet and the nebular symmetry axis, and their location inside or
outside the bright shell of the PNe which is expected to define the
fast vs. slow wind interaction region, are also given in Table~2.

As already stated, a basic prediction of the ISW models, 
within the single-star scenario, is that the
jets have to be roughly coeval with the main nebula. This is true for
four objects in Table~2 (NGC~7009, NGC~6891, NGC~6884 and NGC
3918). Three others (M~1-16, Fg~1 and K~4-47) are older than the
main nebulae and would be better understood within the accretion-disk
models. But the remaining three objects for which this information is
available have smaller kinematical ages than their PNe and cannot be
accounted for by any of these models.

In addition, it is clear that, at variance with the predictions of the 
ISW models, only two jets (over eight objects with adequate data) lie 
along the polar axis of the main nebula. When the
jet--nebula axis angle is relatively small, either stellar or disk 
precession might explain the observed orientation of the jet. However, 
for very 
large angles (from 50\gr\ in NGC~6884 to the extreme ``equatorial'' jets of
Hb~4 and possibly NGC~6210), the kind of extreme disk precession and wobbling
described by Livio \& Pringle (1997) would be needed, but that is expected to
occur only during a very short phase of the disk evolution (see Cox et al. 
2000 and Sahai et al. 2000, on the direct evidence for highly-collimated jets 
close to the equatorial plane in the protoplanetary nebulae CRL2688 and 
Frosty Leo, respectively).

For all but one of the objects in Table~2 for which detailed kinematical
information is available (NGC~3918, He~2-186, K~4-47, Fg~1, M~1-16,
NGC~6543 and NGC~6891) the jets show a roughly linear increase in 
expansion velocity with distance in the radial direction, as if they
consisted of material ejected by the central star with a range of velocities
during a relatively short burst.  Among the models discussed above,
only the MHD ISW models (Garc\'\i a-Segura et al. 1999) predict this
velocity behavior, favoring the latter ones compared to the pure HD
models.  The location outside the nebular rims of all the jets in
Table 2, instead, does not help in distinguishing between the
different models.

In summary, it appears that the single-star HD and MHD ISW models are 
able to explain only some of the observed jets, the latter models being 
favored by the observed roughly linear increase of velocities with distance 
from the central stars. Three objects are instead better understood as 
accretion-disk jets, but still almost one half of the 
sample shows unexpected properties, in terms of ages or jet--nebulae axis
angle. These objects are a real challenge to the ideas on jet
formation and clearly deserve future theoretical and observational
studies.
 
\subsection{Jet-like features}

The list of jet-like LISs is presented in Table~3. For three objects,
the available morphological and kinematical data do not allow us to
disentangle projection effects and determine the basic properties of
the main nebulae and jets (orientation and deprojected expansion
velocities), and we will not discuss these objects further. But for
six PNe (IC~4593, He~2-429, NGC~6881, K~1-2, Wray~17-1, and likely
NGC~6751), observations have shown that their jet-like LIS really
expand with velocities similar, or at least not significantly larger,
than those of the ambient gas that forms the large-scale shells of the
nebulae. This is rather embarrassing, because all models for the
formation of highly collimated structures predict high-velocity
systems\footnote{Note that the ``low-velocity jets'' described in
Soker (1992) are expected to disappear before the formation of the PN
shell.}. In fact, in spite of the unexpectedly low velocities,
jet-like LISs present many similarities with real jets --- they are
commonly two-sided highly collimated structures, radially or
point-symmetrically distributed.

It is interesting to ask if jet-like LIS also present an increasing velocity
outwards, as the real jets do. There are only two cases where adequate 
information exist: K~1-2 and Wray~17-1 (Corradi et al. 1999). In the case of
K~1-2, the velocity increases linearly as the collimated and knotty structure
protrudes from the edge of the main nebular shell. From there, some
acceleration is expected because of the lower density of the outer ambient
material. Thus, the peculiar behavior of the LIS's velocity in the external
regions could just be a consequence of the density stratification of the
ambient gas, and not related to the LIS formation process itself. The
jet-like LIS of Wray~17-1, on the other hand, displays low radial velocities
through almost the entire structure except for its outermost tail. This
suggests that the tail might not be physically related to the innermost parts
of the LIS, but  is instead most probably an ionization effect, i.e., a
lower-ionization region shielded from energetic photons from the central star
by the inner, higher-ionization patch. The large velocities found at the tail
would just represent the general motions of the outer regions of the PN.

The two cases above are complex and warn us of the danger of
simplistic and general explanations. The fact that the other jet-like
LISs are all found outside the rims and have low velocities and high
collimation is puzzling. None of their characteristics gives us clear
indications about the processes that could collimate these 
systems. Jet-like LISs are  hardly understood at present.

\section{Pairs of knots}

\subsection{Theories}

Jet formation models can also be applied to low-ionization knots which
appear in pairs, because knots and filaments along the jets and at
their tips can develop during the evolution of the jet itself, as a
consequence of kink, Vishniac, Rayleigh--Taylor or Kelvin--Helmholtz
instabilities (c.f.  Garc\'\i a-Segura et al. 1999). One may ask why
we do not see the whole jets in such systems. There are two obvious
possibilities: the jet does not exist any longer, or has properties
which make it difficult to observe. In fact, Soker (1990) proposed
that jets formed by disks are the prime cause of pairs of knots, and
showed that, under certain conditions, the jets can be seen only in
the early stages of the PN evolution, but have cool heads which appear
as more permanent features. In a different context, Garc\'\i a-Segura
(1997) argued that a weak radiative cooling in the post-shock gas,
imposed by the wind conditions, also results in the formation of pairs
of knots instead of extended jets.

A different model to explain the formation of symmetrical knots is
that they arise in the concave section of bow-shocks -- i.e. knots
would develop in the stagnation zones of partially collimated stellar
winds (as cool dense portions of gas accumulates in these regions)
aligned with the symmetry axis and attached to the shell (Steffen \&
L\'opez 1998, 2000). In this case high-velocity pairs of knots
directed along the polar axis of the asymmetrical shell are expected.

Finally, it is important to note that all high-velocity knots (FLIERs)
imaged by the {\it HST} (Balick et al. 1998) do not show the outward-facing
bow-shocks expected to develop due to the supersonic velocity of the
knots, in apparent contradiction with all the above models.  However,
Soker \& Reveg (1998) argue that bow-shocks are not expected when the
density contrast between the knot and the ambient medium is moderate,
because instabilities which occur at the knots' surface may
considerably change their geometry, and form extreme outward-pointing
tail structures instead of bow-shocks (see also Jones, Kang \&
Tregillis 1994).

\subsection{Observed properties}

PNe which present symmetrical pairs of low-ionization knots are listed
in Table 4. Note that some objects are also in Table 2 or Table 3,
because some PNe show more than one type of LIS (see fourth column of
Table 1). Hb 4, for instance, has a jet-like pair as well as a pair of
knots (Corradi et al. 1996; Hajian et al. 1997; L\'opez, Steffen \&
Meaburn 1997).

The third column of the Table~4 indicates whether the knots have
peculiar velocities with respect to the ambient gas. Note that nine
PNe in this table possess high-velocity knots, like the well studied
FLIERs (Balick et al. 1998), but only five of them are clearly in the
polar direction. These two properties would indicate that the pairs of
knots in NGC~7009, Hu~2-1, NGC~6905, NGC~6826 and Kj~Pn~8 could be
either the leading front of unobserved jets or produced at the wind's
stagnation zones.

Six PNe in Table~4 (IC~4593, He~1-1, NGC~2440, Wray~17-1, IC~2553 and
NGC~5189) contain pairs of knots with velocities similar to those of
the ambient gas, or even, in three well observed cases, the LISs have
expansion velocities lower than that of the surrounding nebula.  Both
the stagnation zone model and the jet models predict high expansion
velocities, so they fail to explain these low-velocity knots, unless
they have been completely slowed down by interaction with the
surrounding gas\footnote{Severe slowing down is expected when the jet
is not feeding the knot anymore, and it is mainly caused by the
decrease of the knot's density, that follows its expansion once the
knot becomes ionized (Soker \& Reveg 1998).}. Moreover, the high degree
of symmetry of these low-velocity knots also excludes their formation
via in situ instabilities. We are then tempted to suggest that they
could represent symmetric knots originating in the AGB wind, which
survived till the present age. This would imply a peculiar geometry
for the AGB wind. This is an interesting possibility for some objects
(e.g. IC 2553, Corradi et al. 2000b; and the others for which the
low-velocity pairs are located outside the rims, i.e. sharing the
outer shells expansion), but is unlikely to work for the highly
peculiar structures found for instance in Wray 17-1 (cf. Corradi et
al. 1999).

Summarizing, the controversy over the presence of well 
defined bow-shocks at the knots' surface is 
crucial in order to understand the formation of symmetrical, 
high-velocity pairs of knots. The low-velocity pairs of knots 
deserve a different explanation (unless a significant slowing down 
is invoked), and we argue that in some objects these features might 
be fossil remnants of symmetrical condensation originating in the AGB wind.

\section{Non-symmetrical LISs}

Table 5 lists the known PNe associated with non-symmetrical LISs, i.e.
both isolated knots and systems of non-symmetrical knots or filaments
(the latter sometimes radially aligned with the central star). Some of
these PNe have well studied LISs, like the cometary knots of NGC~7293
(O'Dell \& Burkert 1997; Meaburn et al. 1998) or the high-velocity
knots of MyCn~18 (Bryce et al. 1997; O'Connor et al. 2000).

The formation of most isolated/non-sym\-metrical LISs can be understood by
assuming different in situ dynamical and/or radiative
instabilities and other processes that can create inhomogeneities in
the mass outflow, such as condensations created during the
pre-PN stage. The two types of processes -- i) Rayleigh--Taylor,
Kelvin--Hemholtz and Vishniac instabilities (Dyson, Hartquist \& Biro
1993; Garc\'\i a-Segura \& Franco 1996), and ii) fossil AGB
condensations -- would produce structures which do not show highly
peculiar velocities as compared to those of the shells in which they
are embedded.  This might be the case for IC~4593, Hu~2-1, NGC~7662,
NGC~5882 and NGC~6337.

In addition, if resulting from dynamical instabilities during the fast
vs. slow winds interactions, the structures will appear at the edges,
or departing from, the bright rims of the PNe.  Only three PNe in
Table 5 show this kind of LIS (NGC~7293, NGC~6326 and NGC~6337),
whereas most objects (14 over 17) appear well {\it outside} the rims,
often associated with the outer shells (IC~4593, NGC~7662, NGC~2392,
NGC~3242, NGC~5882, and probably NGC~6818, NGC~7354 and IC~4637).  It
is nowadays believed (Marten \& Sch\"onberner 1991; Mellema 1994) that
the dynamical structure of the attached shells of PNe is driven by the
expansion of the ionization front. Thus, the above facts would
indicate that the majority of the isolated LISs are not produced by
dynamical instabilities related to the action of the fast post-AGB
wind.

Five PNe in Table~5, however, possess high-velocity structures. This
raises a problem when trying to explain the formation of these LIS as
above.  Radiation instabilities can modify/\-exacerbate such
structures, and the ionization process itself can increase the LIS
velocity substantially (the ``rocket effect"; Mellema et
al. 1998). So, these LISs might be structures previously formed that
were later accelerated by the rocket effect.  Again, while this might
apply to some of the PNe in Table~5 (NGC~3242 and NGC~2392 being the
best cases), it is very unlikely that it can for instance explain the
very high velocities ($\le$ 500 km s$^{-1}$) of the system of knots of
MyCn~18 (Bryce et al. 1997), which furthermore show the same linear
increase of velocities with distance as several jets discussed in
Section 4.2.

Other models proposed to explain the formation of non-symmetrical LISs
invoke in situ instabilities caused by the interaction of the
expanding AGB wind with a non-homogeneous medium. In fact, Dgani \&
Soker (1997, 1998) and Soker \& Zucker (1997) discussed the
interaction of the PN with the magnetized interstellar medium, showing
that Rayleigh--Taylor and Kelvin--Hemholtz instabilities could fragment
the outermost shell (the halo) producing holes, which in turn allow
the penetration of interstellar material into the internal regions of
the nebula. They argue that such interstellar clumps would appear as
low-ionization knots, and if this is the case, we would expect
velocities generally smaller than that of the shell in which they are
embedded. Thus, noting that only one of the PNe in Table~5 has
isolated LISs with lower expansion velocities than the main nebula,
interstellar clumps are not likely as an explanation for the formation
of isolated structures.

In summary, we find that in situ instabilities and/or fossil
condensations due to an inhomogeneous AGB wind are the primary origin
of isolated LISs. Most of them are located in the outer shells of the
PNe and therefore are not related to the action of the fast post-AGB
wind, but to the dynamical and radiative evolution of the AGB slow
wind.

\section{On PNe showing multiple pairs of LISs}

Several PNe in Table~1, namely IC~4634, Hb~4, NGC~6309, M~3-1, K~1-2, 
Wray~17-1, IC~2553, NGC~5189 and He~2-141, show pairs of LISs which 
appear to be
 roughly perpendicular to each other. Note that this property appears
in all the classes of LISs discussed in this paper (jets, jet-like
features and pairs of knots), adding another complex piece to the
puzzle of the formation of these low-ionization components of PNe.

One may argue that a combination of some of the models previously
discussed could be at work. For instance, the formation of jets by
accretion-disks (i.e. before the PN main bodies) may be followed by
the ISW evolution, which in turn originate a pair of knots in the
polar direction of the main shell, i.e. not necessarily in the
direction of the jets. The simulations of Blackman et al. (2000),
on the interplay between MHD disk winds vs. MHD stellar winds, result
in a different axis for the jet and the nebula, with sometimes very
large angles (up to 90 degrees) between the collimated systems, as
observed in some real PNe.  Unfortunately, neither the PNe mentioned
at the beginning of this section have sufficiently good data, such as
information on the deprojected orientation of the various collimated
structures, nor are the simulations by Blackman et al. (2000)
sufficiently detailed to allow a formal comparison.  The other
possibility of an extreme precession of the main nebular axis or of
the disk appears unlikely, as mentioned in Section 4.2.

\section{Conclusions}

We have studied the 50 PNe presently known to contain 
low-ionization structures and compared their morphological and kinematical
properties with the predictions of current theoretical models. The main
results of this study are the following.

1 - Low-ionization structures are present indistinctly in all
morphological classes of PNe, indicating that their formation is not
necessarily connected with the processes responsible for the
asphericity of the main morphological components of PNe.

2 - Only a few of the observed low-ionization {\it jets} could be
formed by the HD and MHD interacting stellar winds (NGC~7009, NGC~6891
and NGC~3918), the latter ones being favored by the observed linear
increase of the expansion velocity along the jets.  Other jets (K~4-47, M~1-16
and Fg~1) can be better understood adopting accretion-disk jet
models. No model appears instead to be able to explain the jets
younger than the main PN shells (Hb~4, NGC~6210 and NGC~6543), nor
those with very large jet-nebula angles (Hb~4, NGC~6210 and NGC~6884).

3 - A number of {\it jet-like} structures show velocities that are
similar to those of the environment. We have studied five well
observed nebulae containing this kind of LIS (IC~4593, He~2-429,
NGC~6881, K~1-2 and Wray~17-1) and conclude that none of the existing
models can account for them.

4 - Symmetrical pairs of {\it high-velocity knots} could  originate by HD
or MHD interacting stellar winds and accretion-disk systems, or at the
zones of stagnation of partially collimated winds. Since the related 
outward-facing bow shocks are generally not observed, their surfaces 
might have been modified by HD instabilities. Pairs of {\it low-velocity 
knots} are in turn less well studied theoretically, and we suggest that some of 
them might have originated in the AGB wind, implying a very interesting  
AGB mass-loss geometry.

5 - Isolated LISs may be formed by in situ instabilities as well as by
fossil AGB mass loss inhomogeneities. Distinguishing between these
processes is not an easy task; however, the position of most 
isolated LISs indicate that they are not related to dynamical
instabilities due to the action of the fast post-AGB wind.

Clearly, further modeling and observations are needed in order to
reach a more complete understanding of the physical processes
governing the formation of the various types of LISs. In this
direction, one of the next steps will be to try a detailed comparison of the
physico-chemical properties of LISs with those of the main nebular
components, for a broad sample of PNe.

\acknowledgements We would like to thank the referee, Noam Soker, for  
several useful comments and suggestions which improved our paper. 
The work of DRG is supported by a grant 
from the Brazilian Agency FAPESP (98/7502-0) and that of RLMC and AM  
from the Spanish grant DGES PB97-1435-C02-01.

\newpage 
\begin{deluxetable}{llcll}
\tablenum{1}
%\tablewidth{60pc}
\tablecaption{Planetary nebulae showing low-ionization structures}
\tablehead{
\multicolumn{1}{l}{PN name}& 
\multicolumn{1}{l}{PN G number} &
\multicolumn{1}{l}{PN morphology\tablenotemark{1}} &
\multicolumn{1}{l}{Class of LIS} &
\multicolumn{1}{l}{References\tablenotemark{2}}}
\startdata
IC 4634   & 000.3+12.1  & P & 1 pair of jets, 1 pair of knots & 1, 2, 3 \nl
Cn 1-5    & 002.2 $-$09.4   & E & 1 pair of knots & 2 \nl
NGC 6369  & 002.4 +05.8   & E  & several filaments & 3 \nl
Hb 4      & 003.1 +02.9   & E  & 1 pair of jets, 1 pair of knots  & 2, 3, 4 \nl
NGC 6309  & 009.6 +14.8   & P & 1 jet-like pair, 1 pair of knots & 2 \nl
IC 4593   & 025.3 +40.8  & E & 1 jet-like pair, 1 pair of knots and & 2, 5, 6 \nl
          &               &    & other knots &  \nl
NGC 6818  & 025.8 $-$17.9  & E  & several knots & 2 \nl       
NGC 6751  & 029.2 $-$05.9  & E & 1 jet-like pair & 2, 7, 8 \nl
PC 19     & 032.1 +07.0  & P  & 1 pair filaments & 9 \nl 
NGC 7293  & 036.2 $-$57.1  & E  & several knots & 10, 11 \nl
NGC 6572  & 034.6 +11.8  & E & 2 pairs of knots & 12 \nl
NGC 7009  & 037.7 $-$34.5  & E  & 1 pair of jets, 1 pair of knots & 13, 14, 15 \nl
NGC 6210  & 043.1 +37.7  & P  & 1 pair of jets & 16 \nl
He 2-429  & 048.7 +01.9  & E  & 1 jet-like pair & 9 \nl
Hu 2-1    & 051.4 +09.6  & B  & 1 pair knots, 1 isolated knot & 17 \nl
NGC 6891  & 054.1 $-$12.1  & E  & 1 pair of jets& 18 \nl
He 1-1    & 055.3 +02.7   & P  & 1 pair of knots & 9 \nl
K 3-35    & 056.0 +02.0   & E  & 1 pair of knots & 19 \nl
NGC 6905  & 061.4 $-$09.5  & I  & 1 pair of knots & 20, 21 \nl
NGC 6881  & 074.5 +02.1  & B & 2 jet-like pairs, 1 pair of knots & 22 \nl
NGC 6884  & 082.1 +07.0  & E  &  1 pair of jets, 1 pair of knots & 23 \nl
NGC 6826  & 083.5 +12.7  & E  & 1 pair of knots & 14, 15, 24 \nl
NGC 6543  & 096.4 +29.9  & E  & 1 pair of jets & 14, 25 \nl
NGC 7662  & 106.5 $-$17.6 & E  & multiple pairs of knots and filaments and & 15, 26 \nl
          &               &    & some isolated knots & \nl
NGC 7354  & 107.8 +02.3 & E  & 1 jet-like pair, isolated knots & 3 \nl 
Kj Pn 8   & 112.5 $-$00.1 & B & 2 pairs of knots & 27, 28, 29 \nl
K 4-47    & 149.0 +04.4 & -  & 1 pair of jets & 30 \nl
IC 2149   & 166.1 +10.4 & -  & 1 knot & 26 \nl
J 320     & 190.3 $-$17.7 & P & 1 knot & 31 \nl
NGC 2392  & 197.8 +17.4 & E  & several filaments & 31, 32, 33, 34 \nl
M 1-16    & 226.8 +05.6 & B & 1 pair of jets & 35, 36 \nl
NGC 2438  & 231.8 +04.1 & E & 1 pair of knots & 2 \nl
NGC 2440  & 234.8 +02.4 & B  & 1 pair of knots & 37 \nl
M 3-1     & 242.6 $-$11.1 & P  & 1 jet-like pair, 1 pair of knots & 2 \nl
NGC 2452  & 243.3 $-$01.0 & I  & 1 knot, 1 filament & 2 \nl
K 1-2     & 253.0 +10.1 & E  & 1 jet-like pair, 2 pairs of knots & 2, 38 \nl
Wray 17-1 & 258.0 $-$15.7 & E  & 1 jet-like pair, 1 pair of knots & 2, 38 \nl
NGC 3242  & 261.0 +32.0 & E & 2 knots & 2, 15, 26 \nl
IC 2553   & 285.4 $-$05.3 & E  & 2 pairs of knots & 2, 39 \nl
Fg 1      & 290.5 +07.9 & P  & 1 pair of jets & 40, 41, 42 \nl
NGC 3918  & 294.6 +04.7 & E  & 1 pair of jets & 2, 38 \nl
NGC 5189  & 307.2 $-$03.4 & B  & multiple pairs of knots & 24, 43 \nl
MyCn 18   & 307.0 $-$04.1 & B & several knots & 44, 45 \nl
He 2-434  & 320.3 $-$28.8 & E  & 1 pair of knots & 2 \nl
He 2-141  & 325.0 $-$04.1 & I  & 2 pairs of knots & 2 \nl
NGC 5882  & 327.8 +10.0 & E & 3 knots & 39 \nl
He 2-186  & 336.0 $-$05.1 & P  & 1 pair of jets & 2, 30 \nl
NGC 6326  & 338.2 $-$08.3 & E  & several filaments & 2 \nl
IC 4637   & 345.4 +00.1 & E  & 2 knots & 2 \nl
NGC 6337  & 349.3 $-$01.1 & E & several filaments & 2, 30 \nl
\enddata
\tablenotetext{1}{E= ellipticals; B= bipolars or 
quadrupolars; I= irregulars; and P= point-symmetric PNe.}
\tablenotetext{2}{
1= Schwarz 1993; 
2= Corradi et al. 1996; 
3= Hajian et al. 1997; 
4= L\'opez, Steffen \& Meaburn 1997; 
5= Corradi et al. 1997; 
6= O`Connor et al. 1999; 
7= Gieseking \& Solf 1986; 
8= Chu et al. 1991; 
9= Guerrero, V\'azquez \&  L\'opez 1999;  
10= O`Dell \& Burkert 1997; 
11= Meaburn et al. 1998; 
12= Miranda et al. 1999;  
13= Reay \& Atherton= 1985; 
14= Balick et al. 1994; 
15= Balick et al. 1998;  
16= Phillips \& Cuesta 1996; 
17= Miranda 1995; 
18= Guerrero et al. 2000; 
19= Miranda et al. 1998;  
20= Cuesta, Phillips \& Mampaso 1990; 
21= Cuesta, Phillips \& Mampaso 1993; 
22= Guerrero \& Manchado 1998; 
23= Miranda, Guerrero \& Torrelles 1999; 
24= Phillips \& Reay 1983; 
25= Miranda \& Solf 1992; 
26= Balick et al. 1993; 
27= L\'opez, V\'azquez \& Rodr\'\i guez 1995; 
28= L\'opez et al. 1997; 
29= V\'azquez, Kingsburgh \& L\'opez 1998; 
30= Corradi et al. 2000a; 
31= Balick 1987; 
32= Miranda \& Solf 1990; 
33= O`Dell, Weiner \& Chu 1990; 
34= Phillips \& Cuesta 1999; 
35= Schwarz 1992; 
36= Corradi \&  Schwarz 1993; 
37= L\'opez et al. 1998; 
38= Corradi et al. 1999; 
39= Corradi et al. 2000b; 
40= L\'opez, Roth \& Tapia 1993; 
41= L\'opez, Meaburn \& Palmer 1993; 
42= Palmer et al. 1996; 
43= Reay, Atherton \& Taylor 1984; 
44= Bryce et al. 1997; 
45= O'Connor et al. 2000.
}

\tablenotetext{}{Note that IC~4673 and IC~1297, originally
present in the list of LISs of Corradi et al. (1996), are not included in the 
table since recent spectra taken by us removed the original suggestion that
they contain low-ionization microstructures. In particular, the bright knot
found by Corradi et al. (1996) inside IC~1297 turned out to be field star.}
\end{deluxetable}

\newpage 
\begin{deluxetable}{lcccc}
\tablenum{2}
\tablewidth{30pc}
\tablecaption{Planetary nebulae with low-ionization jets}
\tablehead{
\multicolumn{1}{l}{Object} & 
\multicolumn{1}{l}{Confidence\tablenotemark{1}} &
\multicolumn{1}{l}{Kinematical ages\tablenotemark{2}} &
\multicolumn{1}{l}{Orientation\tablenotemark{3}} &
\multicolumn{1}{l}{Location\tablenotemark{4}}}
\startdata
IC 4634   & L & - & -            & outside \nl
Hb 4      & H & younger & 90\gr  & outside \nl
NGC 7009  & L & coeval & 30\gr  & outside \nl
NGC 6210  & L & younger & 90\gr  & outside \nl
NGC 6891  & H & coeval & 0\gr & outside \nl
NGC 6884  & L & coeval & 50\gr \tablenotemark{*} & outside \nl
NGC 6543  & H & younger & 20\gr \tablenotemark{*} & outside \nl
K 4-47    & H & older & - & outside \nl
M 1-16    & L & older & - & outside \nl
Fg 1      & H & older & 40\gr \tablenotemark{*} & outside \nl
NGC 3918  & H & coeval & 0\gr & outside \nl
He 2-186  & H & - & - & outside \nl
\enddata
\tablenotetext{1}{H stands for high and L for low confidence.}
\tablenotetext{2}{Kinematical ages are quoted with respect to the main shell 
ones.
%So that younger (older) means that jets are younger (older) than the 
%main shell. If they have approximately the same age we use coeval.
}
\tablenotetext{3}{The approximate angle of the jet with respect to the major 
axis of the main shell, from polar (0\gr) to equatorial (90\gr) jet 
orientations. `-' is used for the PNe without such a information. `*' 
indicates that the quoted angle is for the jet precessing axis.}  
\tablenotetext{4}{The apparent location of the tip of the jet with respect 
to the rim, or to the barely resolved core emission in the case of  K~4-47.}
\end{deluxetable}

\newpage
\begin{deluxetable}{lcccc}
\tablenum{3}
\tablewidth{30pc}
\tablecaption{Planetary nebulae with jet-like structures}
\tablehead{
\multicolumn{1}{l}{PN name}& 
\multicolumn{1}{l}{Confidence\tablenotemark{1}} &
\multicolumn{1}{l}{Kinematical data\tablenotemark{2}} &
\multicolumn{1}{l}{Orientation\tablenotemark{3}} &
\multicolumn{1}{l}{Location\tablenotemark{4}}}
\startdata
NGC 6309  & L & no  & -          & outside \nl
IC 4593\tablenotemark{a}   & H & yes & -          & outside \nl
NGC 6751  & L & yes & -          & outside \nl
He 2-429  & H & yes &  0\gr & outside \nl
NGC 6881  & H & yes &  40\gr\tablenotemark{*} & outside \nl
NGC 7354  & L & no  & -          & outside \nl
M 3-1     & L & no  & -          & outside \nl
K 1-2     & H & yes & - & inside \nl 
Wray 17-1 & H & yes & -          & inside \nl
\enddata
\tablenotetext{1}{As in Table~2.}
\tablenotetext{2}{Some of the PNe classified as jet-like LISs do not have 
kinematical studies for the LISs (those quoted with `no'). The others are real 
low-velocity systems highly collimated.}
\tablenotetext{3}{As in Table~2, but for the jet-like axis.}
\tablenotetext{4}{As in Table~2, but for the jet-like LISs.}
\tablenotetext{a}{Harrington \& Borkowsky (2000) claim that this highly 
collimated LIS is a real jet, based on {\it HST} imaging. Corradi et al. (1997), 
however, argued that it is unlikely that the very low radial velocities 
observed in the LIS are just a projection effect.}
\end{deluxetable}

\newpage
\begin{deluxetable}{lcccc}
\tablenum{4}
\tablewidth{30pc}
\tablecaption{Planetary nebulae with pairs of low-ionization knots}
\tablehead{
\multicolumn{1}{l}{PN name}& 
\multicolumn{1}{l}{Confidence\tablenotemark{1}} &
\multicolumn{1}{l}{Peculiar velocities\tablenotemark{2}} &
\multicolumn{1}{l}{Orientation\tablenotemark{3}} &
\multicolumn{1}{l}{Location\tablenotemark{4}} 
}
\startdata
IC 4634   & L & - & - & outside \nl
Cn 1-5    & L & - & - & outside \nl
Hb 4      & L & - & - & inside \nl
NGC 6309  & L & - & - & inside \nl
IC 4593   & H & no & - & outside \nl
PC 19     & H & high & - & outside \nl 
NGC 6572  & H & high/low & - & outside \nl %
NGC 7009  & H & high &  polar & outside \nl
Hu 2-1    & H & high &  polar & outside \nl 
He 1-1    & L & no & - & outside \nl 
K 3-35    & H & -  & - & inside \nl 
NGC 6905  & L & high &  polar & outside \nl 
NGC 6881  & H & - & - & outside \nl 
NGC 6884  & L & high & - & outside \nl 
NGC 6826  & L & high &  polar & outside \nl
NGC 7662  & H & high/low & - & outside \nl 
Kj Pn 8   & H & high &  polar & outside \nl 
NGC 2438  & L & - & - & inside \nl 
NGC 2440  & H & no &  polar & outside \nl 
M 3-1     & L & - & - & outside \nl 
K 1-2     & L & - & - & outside \nl
Wray 17-1 & H & low/no & - & inside \nl
IC 2553   & H & no & - & outside \nl 
NGC 5189  & L  & low/no & - & inside \nl 
He 2-434  & L & - & - & outside \nl 
He 2-141  & L & - & - & inside \nl  
\enddata
\tablenotetext{1}{As for Table~2.}
\tablenotetext{2}{High (low) corresponds to peculiar velocities higher (lower) 
than the environment. `no' stands for the cases in which velocities of LIS do 
not differ from the ambient velocity, and `-' for the PNe without kinematical 
data for the LIS. Some PNe have more than one pair of knots with available kinematical 
data. In these case we use `high/low' and `low/no' indicating the kinematical status of 
two pairs of low-ionization knots.}
\tablenotetext{3}{As for Table~2, but for at least one of the pairs of knots.}
\tablenotetext{4}{As for Table~2.}
\end{deluxetable}

\newpage
\begin{deluxetable}{lccc}
\tablenum{5}
\tablewidth{25pc}
\tablecaption{Planetary nebulae with isolated low-ionization structures}
\tablehead{
\multicolumn{1}{l}{PN name}& 
%\multicolumn{1}{l}{PN G} &
\multicolumn{1}{l}{Confidence\tablenotemark{1}} &
\multicolumn{1}{l}{Peculiar velocities\tablenotemark{2}} &
\multicolumn{1}{l}{Location\tablenotemark{3}} 
}
\startdata
NGC 6369  & L & - & outside \nl
IC 4593   & H & no & outside \nl
NGC 6818  & L & - & outside \nl       
NGC 7293  & H & low/high$^*$ & inside \nl
Hu 2-1    & L & no & outside \nl
NGC 7662  & H & no & outside \nl
NGC 7354  & L & - & outside \nl 
IC 2149   & L & high & - \nl
J 320     & L & - & outside \nl
NGC 2392  & H & high & outside \nl
NGC 2452  & L & - & outside \nl
NGC 3242  & H & high & outside \nl
%IC 2553   & H & no & outside \nl
MyCn 18   & H & high & outside \nl
NGC 5882  & H & no & outside \nl
NGC 6326  & L & - & inside \nl
IC 4637   & L & - & outside \nl
NGC 6337  & H & no & inside \nl
\enddata
\tablenotetext{1}{As in Table~2.}
\tablenotetext{2}{As in Table~4, but for isolated LISs.}
\tablenotetext{3}{As in Table~2.}
\end{deluxetable}

\end{document}